\pdfoutput=1

\documentclass[aps,prl,groupedaddress,showpacs,showkeys,twocolumn,amsmath,
amsfonts,amssymb,superscriptaddress]{revtex4-1}

\usepackage{graphicx}
\usepackage[utf8x]{inputenc}
\usepackage{hyperref}
\usepackage{xcolor}
\usepackage{qcircuit}
\usepackage{physics}
\usepackage{amsbsy}
\usepackage{bm}
\newcommand{\Symdiff}{\mathop{\mathchoice{\bigdelta\huge}
                                         {\bigdelta\Large}
                                         {\bigdelta\large}
                                         {\bigdelta\normalsize}}}
\newcommand{\bigdelta}[1]{\vcenter{#1\hbox{\ensuremath{\Delta}}}}
\hypersetup{
    colorlinks,
    linkcolor={red!50!black},
    citecolor={blue!60!black},
    urlcolor={blue!50!black}
}

\begin{document}

\title{Entanglement, quantum correlators and connectivity in graph states}

\author{Arthur Vesperini}
\email[]{a.vesperini@student.unisi.it}
\affiliation{DSFTA, University of Siena, Via Roma 56, 53100 Siena, Italy}
\affiliation{Centre de Physique Th\'eorique, Aix-Marseille University,
Campus de Luminy, Case 907,
13288 Marseille Cedex 09, France}
\author{Roberto Franzosi}
%\email[]{roberto.franzosi@unisi.it}
\affiliation{DSFTA, University of Siena, Via Roma 56, 53100 Siena, Italy}
\affiliation{QSTAR \& CNR - Istituto Nazionale di Ottica,    Largo Enrico Fermi 2, I-50125 Firenze, Italy}
\affiliation{INFN Sezione di Perugia, I-06123 Perugia, Italy}

\date{\today}

\begin{abstract}
In this work, we present a comprehensive exploration of the entanglement and graph connectivity properties of graph states. We quantify the entanglement in pseudo graph states using the entanglement distance, a recently introduced measure of entanglement. Additionally, we propose a novel approach to probe the underlying graph connectivity of genuine graph states, using quantum correlators of Pauli matrices. Our findings also reveal interesting implications for measurement processes, demonstrating the equivalence of certain projective measurements. Finally, we emphasize the simplicity of data analysis within this framework. This work contributes to a deeper understanding of the entanglement and connectivity properties of graph states, offering valuable insights for quantum information processing and quantum computing applications. In this work, we do not resort to the celebrated stabilizer formalism, which is the framework typically preferred for the study of this type of state; on the contrary, our approach is solely based on the concepts of expectation values, quantum correlations and projective measurement, which have the advantage of being very intuitive and fundamental tools of quantum theory.
\end{abstract}

\maketitle

\section{Introduction}

Entanglement, in addition to be one of the most historically puzzling properties of quantum mechanics, constitutes the main resource for quantum cryptography and computation, and quantum-based technologies. The quantum information community developed, in the past decades, an extensive number of approaches to characterize its abundant phenomenology and various properties \cite{guhne_entanglement_2009,horodecki_quantum_2009}.
Entanglement in multipartite states has proven to be a much more involved notion than in bipartite states.

Indeed, as is well-known, entangled bipartite states are such that the measurement of one subsystem \textit{fully determines} the state of the other subsystem, by which we mean that there exists local measurement of which the outcome is certain. 

On the other hand, in entangled multipartite states, while measurements performed on entangled subsystems always, by construction, modifies the rest of the system, it may leave the latter maximally entangled and completely undetermined, by which we mean that there exists no local measurement for which the outcome is certain. 

In this context, understanding and characterizing the connectivity properties of multipartite quantum states arises as an important and complex task. \\

We will hereafter denote $\sigma_k^\mu$, with $k=x,y,z$ the Pauli matrices acting on qubit $\mu$, and abbreviate $\bigotimes\limits_{\nu\in Q}\sigma_k^\nu=\sigma_k^Q$ for any set $Q$ of qubits. We call Pauli observable any operator that can be written as a tensor product of Pauli matrices, i.e. tensor products of traceless hermitian $2\times2$ matrices. \\

Graph states (GS) form a class of maximally entangled pure quantum states that have emerged as a powerful resource for quantum information processing. These states indeed serve as a valuable resource for performing quantum gates and enabling fault-tolerant quantum computing. As such, it constitute the foundation for various quantum computing protocols, in particular the one-way quantum computer, also called measurement-based quantum computer \cite{raussendorf_quantum_2012,hein_multi-party_2004,nielsen_cluster-state_2006,vesperini_quantum_2019,hein_entanglement_2006}.

GS are defined on undirected graphs, where each vertex represents a qubit and edges denote entangling interactions between qubits. In most of the literature, the preferred terminology is to refer to GS defined on lattices as \textit{cluster states}. The following work, however, addresses general GS, hence embedded on arbitrary graphs, not necessarily regular in any way (i.e. not possessing any particular symmetry).\\

Let $G(V,E)$ be an undirected graph with associated set of vertices $V$ and edges $E$. To each vertex $\nu\in V$ corresponds a qubit, while to each edge $(a,b)\in E$ corresponds an interaction represented by some non-local unitary operator of the form 
\begin{equation}
U_{ab}(\varphi_{ab}) = e^{-i\frac{\varphi_{ab}}{4}}e^{i\frac{\varphi_{ab}}{4}\sigma_z^a}e^{i\frac{\varphi_{ab}}{4}\sigma_z^b}e^{-i\frac{\varphi_{ab}}{4}\sigma_z^a\sigma_z^b}
\end{equation}

We will here, for the sake of simplicity, assume an equal strength of interaction for each edge, that is: $\forall (a,b),\;\varphi_{ab}=\varphi$.

Consider the initial product state $|\Psi\rangle=|+\rangle^{V}$, where $|+\rangle=\frac{1}{\sqrt{2}}(|0\rangle+|1\rangle)$ is the eigenstate of $\sigma_x$ of eigenvalue $+1$. We then define the pseudo graph state (PGS) as $|G(\varphi)\rangle = \prod_{(a,b)\in E}U_{ab}(\varphi)|\Psi\rangle$. Genuine GS correspond to the case $\varphi=\pi$.\\

In this work, we start by quantifying the entanglement in the general case of PGS, using the entanglement distance (ED), a measure of entanglement recently introduced in Ref. \cite{cocchiarella_entanglement_2020}. We then explore a novel approach to probe the underlying graph connectivity of genuine GS, using correlators of Pauli matrices. We first do so by studying correlations of pairs of qubits (i.e. \textit{two-point correlators}), revealing how these quantities solely depend on the relation between their respective neighbourhoods (namely, in the language of graph theory, if they are \textit{twins}, \textit{adjacent twins}, leaf vertices, etc...); we then notice the possibility of more general probing of graph properties, through the use of higher order correlators (i.e. involving more than two qubits); we further remark interesting implications in terms of measurement processes, namely how our approach can highlight the equivalence of some projective measurements; we then underline the simplicity of data analysis in this context, coming from the fact that all of the correlators derived take values $-1$, $0$ or $1$. We conclude this work by summing up the advantages of our method, with respect to the stabilizer formalism, and underline how both complete each other in the aim of characterizing GS and use them as building blocks for quantum algorithms.\\

%Doing so, we propose a valuable tool for investigating the intricate structure of these quantum states and shedding light on their potential applications in quantum computing.
%This approach opens up a new perspective for investigating the features of quantum computing systems based on GS, without resorting to the stabilizer formalism, traditionally employed to study such systems.
%By unraveling the underlying graph structure, we can optimize quantum algorithms, design efficient quantum circuits, and explore fault-tolerant quantum computing architectures. Furthermore, this approach can aid in identifying and mitigating potential sources of decoherence and noise that can affect the performance of quantum systems.

\section{Entanglement in pseudo graph states}
The ways of quantifying entanglement in multipartite states are manifold \cite{guhne_entanglement_2009,horodecki_quantum_2009}. In this work, we will solely refer to \textit{qubit-wise entanglement}, that is entanglement of bipartitions $(\mu,\mu^C)$, where $\mu$ is a qubit, and $\mu^C$ is its complement relative to the set of all qubits in the system.
 
The entanglement distance (ED), first defined in Ref. \cite{cocchiarella_entanglement_2020}, is an entanglement measure for general multipartite pure states; it has been adapted in Ref. \cite{vesperini_entanglement_2023} to the more general framework of multipartite mixed states, and it has already found since then some interesting applications \cite{vafafard_multipartite_2022,nourmandipour_entanglement_2021,vesperini_correlations_2023}. It finds its theoretical grounds on the Fubini-Study metric associated to the local-unitary invariant projective Hilbert space, called in this context the Entanglement Metric. 

The single-qubit ED is defined as
\begin{equation}\label{single-qubit_ED}
\begin{split}
E_\mu(\ket{s}):&=\min_{\bm{v}^\mu}\;g_{\mu\mu}\big(\ket{s},\bm{v}^\mu\big)\\
&= 1 - \max_{\bm{v}^\mu}|\bra{s}\sigma_{\bm{v}}^\mu\ket{s}|^2\\
&= 1 - |\bra{s}\bm{\sigma}^\mu\ket{s}|^2,
\end{split}
\end{equation}
which equates $1$ if $\mu$ is maximally entangled with the rest of the system, and $0$ if it is fully factorizable.

We choose here to use the latter definition of entanglement, which possesses the advantage of being very easy to compute, relative to the von Neumann entropy. We further define the total entanglement of a state as $\sum\limits_{\mu\in Q}E_\mu(\ket{s})$.\\

From the anticommutation relations of the Pauli matrices 
\begin{equation}
\{\sigma_i^\mu,\sigma_j^\nu\}=2\mathbb{I}\delta_{ij}\delta_{\mu\nu}+2\sigma_i^\mu\sigma_j^\nu(1-\delta_{\mu\nu}),
\end{equation}
we straightforwardly derive:
\begin{equation}\label{commutgraph}
\begin{split}
        \sigma_x^a U_{ab}(\varphi) &= e^{-i\frac{\varphi}{2}\sigma_z^a}e^{i\frac{\varphi}{2}\sigma_z^a\sigma_z^b}U_{ab}(\varphi)\sigma_x^a \\ 
        \sigma_y^a U_{ab}(\varphi) &= e^{-i\frac{\varphi}{2}\sigma_z^a}e^{i\frac{\varphi}{2}\sigma_z^a\sigma_z^b}U_{ab}(\varphi)\sigma_y^a \\
        \sigma_z^a U_{ab}(\varphi) &= U_{ab}(\varphi)\sigma_z^a \\
         \sigma_j^\nu U_{ab}(\varphi)&= U_{ab}(\varphi)\sigma_j^\nu \;,\;\forall j=x,y,z,\; \forall\nu\neq a,b
\end{split}
\end{equation}

The expectation values of the first Pauli matrix hence write
\begin{equation}
\begin{split}
        \langle G(\varphi)|\sigma_x^\nu|G(\varphi)\rangle &=  \langle\Psi|\Big(\prod_{(a,b)\in E}U_{ab}^\dagger(\varphi)\sigma_x^\nu U_{ab}(\varphi)\Big)|\Psi\rangle \\
        & = \langle\Psi|\Big(\prod_{\mu\in N(\nu)} e^{-i\frac{\varphi}{2}\sigma_z^\nu}e^{i\frac{\varphi}{2}\sigma_z^\nu\sigma_z^\mu}\Big)\sigma_x^\nu|\Psi\rangle \\
        &=  \langle\Psi|e^{-i\frac{n_\nu\varphi}{2}\sigma_z^\nu} \Big(\prod_{\mu\in N(\nu)} e^{i\frac{\varphi}{2}\sigma_z^\nu\sigma_z^\mu}\Big)|\Psi\rangle \\
       % &= \langle\Psi|\big(\cos(\frac{n_\nu\varphi}{2})\mathbb{I} -i\sin(\frac{n_\nu\varphi}{2})\sigma_z^\nu\big) \Big(\prod_{\mu\in N(\nu)}\big(\cos(\frac{\varphi}{2})\mathbb{I} +i\sin(\frac{\varphi}{2})\sigma_z^\nu\sigma_z^\mu\big)\Big)|\Psi\rangle \\
        &= \cos(n_\nu\varphi/2)\cos^{n_\nu}(\varphi/2)\;,
\end{split}
\end{equation}
where $N(\nu)$ is the set of the first neighbours of $\nu$, and  $n_\nu=|N(\nu)|$ is its cardinality. We used the fact that all the terms including a Pauli matrix $\sigma_z^\mu$ acting on some $\mu\in N(\nu)$ vanish, since they appear only once and $\forall\mu,\;\langle\Psi|\sigma_z^\mu|\Psi\rangle=0$. 

The expectation values of the second Pauli matrix write
\begin{equation}
\begin{split}
        \langle G(\varphi)|\sigma_y^\nu|G(\varphi)\rangle &=  \langle\Psi|\Big(\prod_{(a,b)\in E}U_{ab}^\dagger(\varphi)\sigma_y^\nu U_{ab}(\varphi)\Big)|\Psi\rangle \\
        &= -i\langle\Psi|\Big(\prod_{\mu\in N(\nu)} e^{-i\frac{\varphi}{2}\sigma_z^\nu}e^{i\frac{\varphi}{2}\sigma_z^\nu\sigma_z^\mu}\Big)\sigma_y^\nu|\Psi\rangle \\
        &=  -i\langle\Psi|e^{-i\frac{n_\nu\varphi}{2}\sigma_z^\nu} \Big(\prod_{\mu\in N(\nu)} e^{i\frac{\varphi}{2}\sigma_z^\nu\sigma_z^\mu}\Big)|\Psi^-_\nu\rangle \\
      %  &= -i\langle\Psi|\big(\cos(\frac{n_\nu\varphi}{2})\mathbb{I} -i\sin(\frac{n_\nu\varphi}{2})\sigma_z^\nu\big) \Big(\prod_{\mu\in N(\nu)}\big(\cos(\frac{\varphi}{2})\mathbb{I} +i\sin(\frac{\varphi}{2})\sigma_z^\nu\sigma_z^\mu\big)\Big)|\Psi^-_\nu\rangle \\
        &= -\sin(n_\nu\varphi/2)\cos^{n_\nu}(\varphi/2),
\end{split}
\end{equation}%$|\Psi^-_\nu\rangle = |+\rangle^{\otimes\nu-1}\otimes|-\rangle\otimes|+\rangle^{\otimes M-\nu}$
where $|\Psi^-_\nu\rangle = |+\rangle^{V\setminus\{\nu\}}\otimes|-\rangle^\nu$, that is, the pure product state with every qubit in the state $|+\rangle$ except for qubit $\nu$ which is in the state $|-\rangle$. The final result stems from the fact that the only non vanishing terms are the ones including one and only one Pauli matrix $\sigma_z^\nu$ acting on $\nu$, since $\forall\mu,\;\langle\Psi|\sigma_z^\mu|\Psi^-_\nu\rangle=\delta_{\mu\nu}$.

Finally, the commutation relations \eqref{commutgraph} trivially imply
\begin{equation}
\begin{split}
     \langle G(\varphi)|\sigma_z^\nu|G(\varphi)\rangle &	=  \langle\Psi|\Big(\prod_{(a,b)\in E}U_{ab}^\dagger(\varphi)\sigma_z^\nu U_{ab}(\varphi)\Big)|\Psi\rangle \\
     &= \langle\Psi|\sigma_z^\nu|\Psi\rangle = 0
\end{split}
\end{equation}

\begin{figure}[htb!]
    \centering
    \includegraphics[width=1.\linewidth]{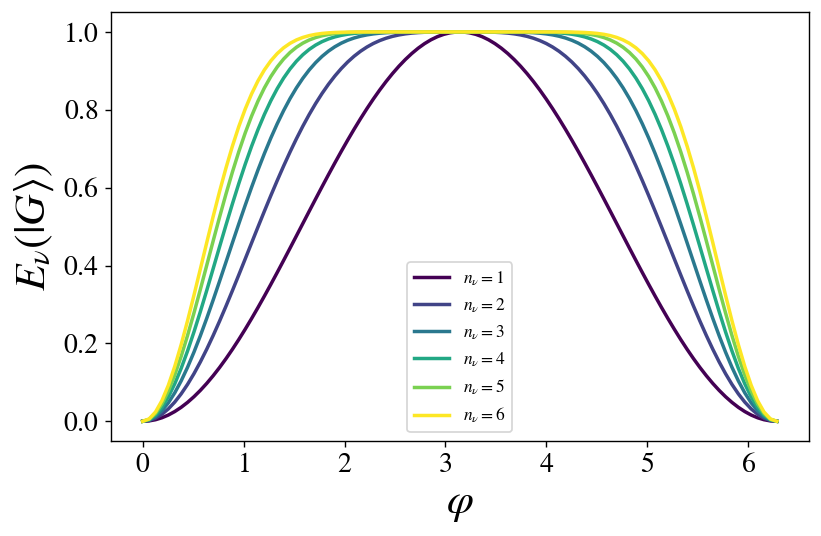}
    \caption{The entanglement distance of a single qubit, as a function of the interaction strength (or duration), for different numbers $n_\nu$ of nearest neighbours. The numerical results agree perfectly with the analytical one of Eq. \eqref{EDgraph}.}
    \label{figEDphimnu}
\end{figure}

\begin{figure}[htb!]
    \centering
    \includegraphics[width=1.\linewidth]{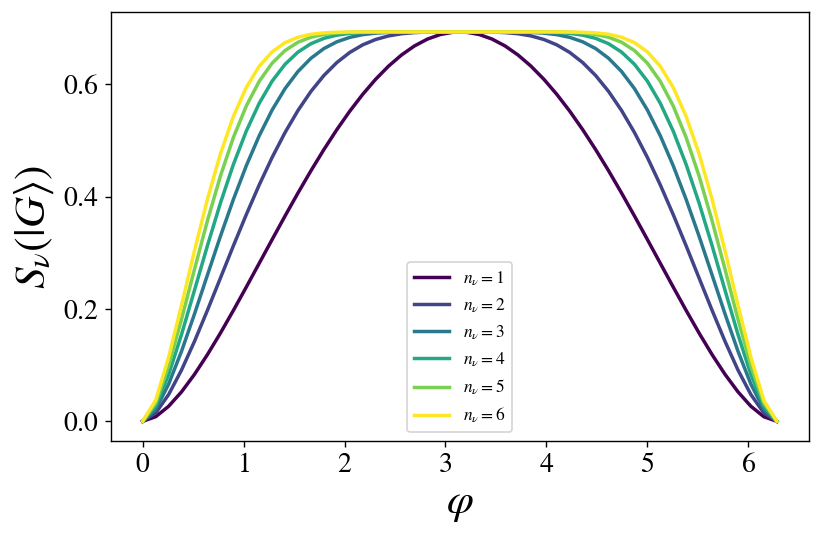}
    \caption{The entropy of entanglement for a bipartition $\nu/\nu^C$, as a function of the interaction strength (or duration), for different numbers $n_\nu$ of nearest neighbours, numerically computed. The scaling and behaviour of this well known measure of bipartite entanglement is evidently very similar to that of the ED.}
    \label{figSphimnu}
\end{figure}

\begin{figure}[htb!]
    \centering
    \includegraphics[width=1.\linewidth]{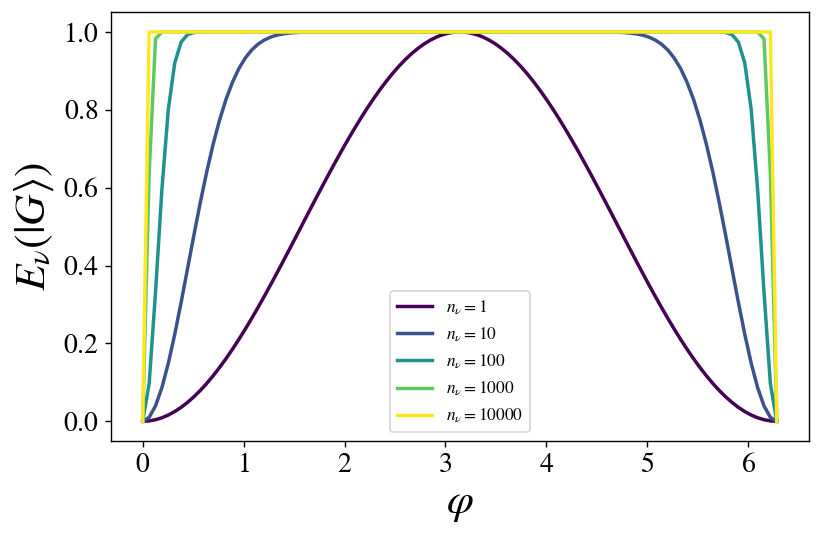}
    \caption{The entanglement distance of a single qubit, as a function of the interaction strength (or duration), for different numbers $n_\nu$ of nearest neighbours.}
    \label{figEDphimnu_pow10}
\end{figure}

It result that the single-qubit ED of a given of a given qubit $\nu$ in a PGS depends on both the interaction strength $\varphi$ and on the number $n_\nu$ of its nearest neighbours
\begin{equation}\label{EDgraph}
    E_\nu(|G(\varphi)\rangle) = 1 - \cos(\varphi/2)^{2n_\nu}
\end{equation}
The numerical confirmation of this result is displayed in Fig. \ref{figEDphimnu}.

As stated before, the value $\varphi=\pi$ corresponds to the genuine GS, in which every non isolated qubit is maximally entangled, regardless of the number of its neighbours. Consider a PGS close to the genuine GS, i.e. where this typical interaction strength is added with a small error $\delta\varphi$, we retrieve
\begin{equation}\label{EDgraph_approx}
    E_\nu(|G(\pi+\delta\varphi)\rangle) \approx 1 - \Big(\frac{\delta\varphi}{2}\Big)^{2n_\nu},
\end{equation}
hence the qubits in a quasi GS get exponentially closer to the maximal value of entanglement as the number of their nearest neighbours increases. The only non trivial case where the small error could be relevant is the one of a qubit with only one link, where the correction is of $o(\delta\varphi^2)$.

It results, as Fig. \ref{figEDphimnu_pow10} emphasizes, that the limit for a large number of bounds writes
\begin{equation} 
E_\nu(|G(\varphi)\rangle) \underset{n_\nu\to\infty}{\longrightarrow} \begin{cases}
0 & \text{if  } \varphi=2n\pi,\,\forall n\in\mathbb{N} \\
1 & \text{else.} \end{cases}
\end{equation}
i.e., up to a null measure set of values of $\varphi$, the ED of a single qubit approaches $1$ when the number of its neighbours becomes very large. In other words, even if the pairwise interaction is very weak, the qubit-wise entanglement, in the sense of \eqref{single-qubit_ED}, can be very close to its maximal value. 

Note that, as can be seen in Fig. \ref{figSphimnu} the entropy of entanglement shows the same behaviour and scaling as the ED, suggesting that the later stems as a valid alternative to the former as a measure of bipartite entanglement. It also has the benefit of being easier to compute, both numerically and analytically, as it only requires the calculation of expectation values, in contrast with the entropy of entanglement, which requires to compute partial trace and matrix logarithms.

\section{Correlators and the effects of measurement in graph states}

We now focus on the case of \textit{genuine} GS, i.e. when $\varphi=\pi$. In particular, we want to compute the off-diagonal elements of the EM, for which we need the various two-point correlators. We denote 
\begin{equation}\label{totalU}
U_{_G}:=\prod_{(a,b)\in E}U_{ab}(\varphi=\pi)=\prod_{(a,b)\in E}\frac{\mathbb{I}+\sigma_z^a+\sigma_z^b-\sigma_z^a\sigma_z^b}{2}.
\end{equation}

From \eqref{commutgraph}, we derive the commutation relations 
\begin{equation}
\begin{split}
& \sigma_x^a U_{_G} = U_{_G} \sigma_z^{N(a)} \sigma_x^a \\
& \sigma_y^a U_{_G} = U_{_G} \sigma_z^{N(a)} \sigma_y^a \\
& \sigma_z^a U_{_G} = U_{_G} \sigma_z^a,
\end{split}
\end{equation}

Note that, for two ensembles $A$ and $B$, we have
$$\sigma_z^A \sigma_z^B = \sigma_z^{A\cup B} = \sigma_z^{A\bigtriangleup B}, $$ where $A\bigtriangleup B = (A\cup B)\setminus (A\cap B)$ is the symmetric difference between sets $A$ and $B$. 

This operation is commutative and associative.
Remark that $A\bigtriangleup B =\emptyset$ if and only if $A=B$.
Furthermore, $\Symdiff\limits_i A_i := A_0\bigtriangleup A_1 \bigtriangleup \cdots \bigtriangleup A_k \bigtriangleup \cdots = \emptyset$ if and only if $\forall \nu$, there is an even  number $k$ of sets $A_i$ containing $\nu$.\\

We can now calculate the correlators, taking advantage of the fact that $\forall A\neq\emptyset,\;\langle\Psi|\sigma_z^A|\Psi\rangle = 0$.

\subsection{Two-points correlators}
We start here by computing pair-wise correlations.
\begin{equation}\label{corr_xx}
\begin{split}
\langle G|\sigma_x^\nu\sigma_x^\mu|G\rangle & =  \langle\Psi|U_{_G}\sigma_x^\nu \sigma_x^\mu U_{_G}|\Psi\rangle \\
& = \langle\Psi|\sigma_z^{N(\nu)}\sigma_z^{N(\mu)}|\Psi\rangle \\
& = \langle\Psi|\sigma_z^{N(\nu)\bigtriangleup N(\mu)}|\Psi\rangle \\
& = \begin{cases}
1 & \text{if } N(\nu)=N(\mu), \\
0 & \text{else.} \end{cases}
\end{split}
\end{equation}
since $(N(\nu)\cup N(\mu))\setminus (N(\nu)\cap N(\mu))=\emptyset$ if and only if $N(\nu)= N(\mu)$. In terms of graph theory, $\langle G|\sigma_x^\nu\sigma_x^\mu|G\rangle=1$ if and only if $\mu$ and $\nu$ are \textit{twins}.

\begin{equation}
\begin{split}
\langle G|\sigma_x^\nu\sigma_y^\mu|G\rangle & = \langle\Psi|U_{_G}\sigma_x^\nu(-i\sigma_z^\mu\sigma_x^\mu)U_{_G}|\Psi\rangle \\
& = -i\langle\Psi|\sigma_z^{N(\nu)}\sigma_z^{N(\mu)}\sigma_z^\mu|\Psi\rangle \\
& = -i\langle\Psi|\sigma_z^{N(\nu)\bigtriangleup N(\mu)\bigtriangleup\{\mu\}}|\Psi\rangle \\
& = 0,
\end{split}
\end{equation}
because, the graph being undirected, if $\nu\in N(\mu)$ then also $\mu\in N(\nu)$, hence $N(\nu)\bigtriangleup N(\mu)\neq\{\mu\}$.

\begin{equation}
\begin{split}
\langle G|\sigma_x^\nu\sigma_z^\mu|G\rangle &= \langle\Psi|U_{_G}\sigma_x^\nu \sigma_z^\mu U_{_G}|\Psi\rangle\\
& = \langle\Psi|\sigma_z^{N(\nu)}\sigma_z^\mu|\Psi\rangle \\
& = \langle\Psi|\sigma_z^{N(\nu)\bigtriangleup\{\mu\}}|\Psi\rangle \\
& = \begin{cases}
1 & \text{if } N(\nu)=\{\mu\}, \\
0 & \text{else.} \end{cases}
\end{split}
\end{equation}
In terms of graph theory, $\langle G|\sigma_x^\nu\sigma_z^\mu|G\rangle=1$ if and only if $\nu$ is a \textit{leaf vertex} (or \textit{pendant vertex}) attached to $G$ through $\mu$.

\begin{equation}
\begin{split}
\langle G|\sigma_y^\nu\sigma_y^\mu|G\rangle & = \langle\Psi|U_{_G}(i\sigma_x^\nu\sigma_z^\nu)(-i\sigma_z^\mu\sigma_x^\mu) U_{_G}|\Psi\rangle \\
& = \langle\Psi|\sigma_z^{N(\nu)}\sigma_z^\nu\sigma_z^\mu\sigma_z^{N(\mu)}|\Psi\rangle \\
& = \langle\Psi|\sigma_z^{N(\nu)\bigtriangleup\{\nu\}\bigtriangleup N(\mu)\bigtriangleup\{\mu\}}|\Psi\rangle \\
& = \langle\Psi|\sigma_z^{\big(N(\nu)\cup\{\nu\}\big)\bigtriangleup \big(N(\mu)\cup\{\mu\}\big)}|\Psi\rangle \\
& = \begin{cases}
1 & \text{if } N(\nu)\cup\{\nu\}=N(\mu)\cup\{\mu\}, \\
0 & \text{else.} \end{cases}
\end{split}
\end{equation}
In terms of graph theory, $\langle G|\sigma_y^\nu\sigma_y^\mu|G\rangle=1$ if and only if $\mu$ and $\nu$ are \textit{adjacent twins}.

\begin{equation}
\begin{split}
\langle G|\sigma_y^\nu\sigma_z^\mu|G\rangle
&= \langle\Psi|U_{_G}(i\sigma_x^\nu\sigma_z^\nu)\sigma_z^\mu U_{_G}|\Psi\rangle \\
& = i\langle\Psi|\sigma_z^{N(\nu)}\sigma_z^\nu\sigma_z^\mu|\Psi\rangle \\
& = i\langle\Psi|\sigma_z^{\big(N(\nu)\cup\{\nu\}\big)\bigtriangleup\{\mu\}}|\Psi\rangle \\
& = 0,
\end{split}
\end{equation}

\begin{equation}
\langle G|\sigma_z^\nu\sigma_z^\mu|G\rangle = \langle \Psi|\sigma_z^\nu\sigma_z^\mu|\Psi\rangle = 0,
\end{equation}

For two arbitrary measurements, determined by the unitary vectors $\bm{v}^\nu$ and $\bm{v}^\mu$, the correlation then writes
\begin{equation}
\begin{split}
\langle G|\sigma_{\bm{v}}^\nu\sigma_{\bm{v}}^\mu|G\rangle & = \sum_{i,j=x,y,z} v_i^\nu v_j^\mu\langle G|\sigma_i^\nu\sigma_j^\mu|G\rangle \\
 = \; & v_x^\nu v_x^\mu \;\text{ if } N(\nu)=N(\mu) \\
 + &v_x^\nu v_z^\mu \;\text{ if } N(\nu)=\{\mu\}\\
 + &v_z^\nu v_x^\mu \;\text{ if } N(\mu)=\{\nu\} \\
 + &v_y^\nu v_y^\mu \;\text{ if } N(\nu)\cup\{\nu\}=N(\mu)\cup\{\mu\}.
\end{split}
\end{equation}
It is fairly obvious that any such correlator can henceforth be fully determined by a quick inspection of the adjacency matrix $A_G$ associated to $G$. For instance, the condition $N(\nu)=N(\mu)$ is equivalent to $A_\nu=A_\mu$

This result makes it clear that non-vanishing pair-wise correlations arise only for very specific connectivity properties of the sites being considered. More precisely, graphs which contain neither twins, nor adjacent twins, nor leaf vertex, have only vanishing pair-wise correlations. This is for instance the case for regular lattices.

Quite interestingly, this also implies that, in GS, most measurements that can be performed on one qubit yields no information on other qubits, and leaves the rest of the system entangled. Such entangled states hence contain \textit{persistent entanglement}: a relatively big number of measurements are necessary to completely break their entanglement.\\

One can also exploit the properties of these correlators to probe the connectivity properties of a graph. Such a procedure could be for instance useful to check that, in a physical apparatus realizing the GS, the implementation of the link operators $U_{ab}$ was successful and free of errors (that would be, the unwanted presence or absence of some of them).

From the above results, checking for twins, adjacent twins and leaf vertices will follow a fairly obvious measurement procedure.
Yet it is possible to go further and check for instance for mere pair-wise neighbourhood, by removing irrelevant vertices from the graph.
To do this, we can use the well-known fact that projective measurement of a single qubit in the direction $z$ effectively removes it from the graph, i.e. isolates it (ref). Formally,
\begin{equation}
\begin{split}
P_{z\pm}^a|G\rangle &= P_{z\pm}^a U_{_G}|\Psi\rangle = U_{_G} P_{z\pm}^a|+\rangle^a\otimes|+\rangle^{V\setminus a}\\
&=\begin{cases} 
&\frac{1}{\sqrt{2}}U_{_G} |0\rangle^a\otimes|+\rangle^{V\setminus a} = \frac{1}{\sqrt{2}}|0\rangle^a\otimes|G\setminus a\rangle \\
&\frac{1}{\sqrt{2}}U_{_G} |1\rangle^a\otimes|+\rangle^{V\setminus a} = \frac{1}{\sqrt{2}}|1\rangle^a\otimes\sigma_z^{N(a)}|G\setminus a\rangle. \end{cases}
\end{split}
\end{equation}
Since $\sigma_z^{N(a)}|G\setminus a\rangle$ is local-unitary equivalent to $|G\setminus a\rangle$, such projective measurement results in an equivalent statistics as the desired GS with graph $G\setminus a$, up to some rotations of the measurement axis.

With a few computation, it can easily be checked that
\begin{equation}
\begin{split}
&\bra{G}\left(\prod_{\mu\neq a,b}P_{z\pm}^\mu\right)\sigma_y^a\sigma_y^b\left(\prod_{\mu\neq a,b}P_{z\pm}^\mu\right)\ket{G} \\
&= \begin{cases} \pm1 & \text{if } b\in N(a) \left(\Leftrightarrow a\in N(b)\right) \\
0 & \text{else.}
\end{cases}
\end{split}
\end{equation}

It is hence enough, in order to examine the existence of a given link $(a,b)$, to perform a projective measurement on the rest of the graph, or at least on the sites that may be linked to $a$ or $b$, prior to measuring the correlator $\braket{\sigma_y^a\sigma_y^b}$. 

This trick can be useful in practice because, in a physical implementation, one may 

\subsection{Higher order correlators}

The inspection of higher order correlators can be used to retrieve informations on more general properties of the graph.

\subsubsection{Neighbourhood probing}
Given an educated guess $\widetilde{N}(\nu)$ for the neighbourhood of $\nu$, one can check its validity by computing the correlator
\begin{equation}\label{neighb_probing}
\langle G|\sigma_x^\nu\sigma_z^{\widetilde{N}(\nu)}|G\rangle\begin{cases} 1 &\text{if  } \widetilde{N}(\nu)=N(\nu) \\ 0 &\text{else,} \end{cases}
\end{equation}

\subsubsection{Topological probing}
The correlator
\begin{equation} \label{fullcorrx}
\langle G|\sigma_x^V|G\rangle = \begin{cases} 1 &\text{if  } \Symdiff_{\mu\in V} N(\mu)=\emptyset \\ 0 &\text{else,} \end{cases}
\end{equation}
results in $1$ if and only if every site has an even number of neighbours.\\

Furthermore,
\begin{equation} \label{fullcorry}
\langle G|\sigma_y^V|G\rangle = \begin{cases} i^{|V|} &\text{if  } \Symdiff_{\mu\in V} \big(N(\mu)\cup \{\mu\}\big)=\emptyset \\ 0 &\text{else,} \end{cases}
\end{equation}
results in $\pm 1$ if and only if every site has an odd number of neighbours. 
It is $1$ if $|V|\bmod 4 = 0$, $-1$ if $|V|\bmod 4 = 2$. 

\textit{Euler's handshaking lemma} states that, in any undirected graph, there is always an even number of vertices $\nu$ such that $n_\nu$ is odd. This guarantees that, as expected, this correlator never takes imaginary values. \\

In particular, if both \eqref{fullcorrx} and \eqref{fullcorry} are null, $G$ is not a regular graph (i.e. for which $\exists k\in\mathbb{N}$ such that $\forall\nu,\;n_\nu=k$). For instance, it can't be a lattice with periodic boundary conditions.

\subsection{Relation to measurement processes}

As already mentioned in the introduction, GS were proposed as a support for measurement-based quantum computation. To this aim, the system is first prepared in a graph state of which the associated graph $G(V,E)$ is a regular lattice (usually, a finite square lattice). Then, a quantum circuit is built from this state by performing series of local projective measurements. 

Hereafter, we thus investigate the effects of such measurements on the overall state, in the light shed by the above results. \\

As noticed in Ref. \cite{vesperini_correlations_2023}, if the expectation value of a product of Pauli observables (i.e. any product of Pauli matrices) on a given pure state $\ket{s}$, i.e. a generalized correlator, equates $1$, then these observables are equivalent with respect to this state. Namely, they act on the state in the same fashion, and the associated projective measurements are themselves equivalent. 

Formally, for any couple of observables $A,\;B$ such that $A^2=B^2=\mathbb{I}$, $\bra{s}AB\ket{s}=1$ implies
\begin{equation}\label{eqv_proj_meas}
\begin{split}
AB\ket{s}&  = \ket{s} \\
 B\ket{s}&  = A\ket{s} \\
 P_B\ket{s}&  = P_A\ket{s} \\
 P_B\ket{s}&  = P_B P_A\ket{s},
\end{split}
\end{equation}
where $P_O=\frac{1}{2}\left(\mathbb{I} + O\right)$ are projectors onto the eigenstates of $O$ of eigenvalue $+1$. 

The projective measurement of $A$ is thus equivalent to that of $B$.\\

For instance, Eq. \eqref{corr_xx} implies that, if $\mu$ and $\nu$ are twin vertices, the projective measure of $\sigma_x^\nu$ is equivalent to that of $\sigma_x^\mu$.\\

The case of higher order correlators leads to somewhat less trivial observations. Consider a measurement of $\sigma_x^\nu$ with an outcome of $+1$. Formally, this corresponds to applying the projector $P_x^\nu=\frac{1}{2}\left(\mathbb{I} + \sigma_x^\nu\right)$ to the graph state $\ket{G}$, up to renormalization. Yet Eq. \eqref{neighb_probing} together with Eq. \eqref{eqv_proj_meas} tell us that this is in fact equivalent to applying $P_z^{N(\nu)}=\frac{1}{2}\left(\mathbb{I} + \sigma_z^{N(\nu)}\right)$. Notice that the latter projector is a non-local one, as it can't be written as the product of local single-qubit projectors; its effect is to project $\ket{G}$ onto the subspace $\left\{\ket{\varphi}\Big|\;\sigma_z^{N(\nu)}\ket{\varphi}=\ket{\varphi}\right\}$. 

Non-locality implies that it does not correspond in itself to any physical measurement process, and rather stems as an entangling operation. It may indeed map a product state to an entangled state.

Let us examine further the effect of this projector on a GS. Omitting the renormalization factor, we obtain
\begin{equation}
\begin{split}
P_x^\nu\ket{G}&=P_z^{N(\nu)}\ket{G}=P_z^{N(\nu)}U_{_G}\ket{\Psi}=U_{_G} P_z^{N(\nu)} \ket{\Psi}\\
&=\frac{1}{2}U_{_G} \ket{+}^{V\setminus N(\nu)}\otimes\left(\ket{+}^{N(\nu)} + \ket{-}^{N(\nu)}\right).
\end{split}
\end{equation}
It results that, as can also be seen by considering the commutation relations \eqref{commutgraph}, the operation $U_{_G}P_x^\nu U_{_G}$ effectively entangles every qubit $\mu\in N(\nu)$ in a state local-unitary equivalent to the Greenberger–Horne–Zeilinger state of $n_\nu$ qubits, a prototypical case of maximally entangled state.

\subsection{Remarks on the simplicity of data analysis}

In an ideal setting, relatively few measurements should, in principle, be enough to compute all of these correlators. \\
This is due to the fact that, for perfect GS, their outcomes can only be $1$, $-1$ or $0$. 
Yet the measurement of a Pauli observable can only result in outcomes of $\pm1$, whether it is a single-qubit or a multi-qubit (i.e. correlator) observable. 

Hence if the statistics yields, for a given Pauli observable $P$, an expectation value of $\langle G|P|G\rangle=1$, we expect to measure only ones. It is thus enough to have measured a single $-1$ to conclude that $\langle G|P|G\rangle=0$. The same reasoning obviously applies to the case of opposite value $\langle G|P|G\rangle=-1$.

Conversely, if the statistics yields an expectation value of $0$, the probability of a measurement outcome $\pm1$ is $\frac{1}{2}$, hence a series of uniform measurement becomes exponentially less likely as the number of measurement $M$ grows.
Precisely, if the value $1$ has been measured $M$ times in a row (and the value $-1$ has never been measured) the statistics yields $\langle G|P|G\rangle=1$ with a probability of $1-2^{-M}$. 
Hence one would need at most $M=-\log_2{(\epsilon)}$ measurement samples to retrieve the true statistics with a confidence of $1-\epsilon$.\\

In a \textit{quasi} GS (i.e. $\varphi=\pi +\delta\varphi$), the link operators write $U_{ab}(\pi+\delta\varphi)=U_{ab}\,\delta U_{ab}$, with
\begin{equation}
\delta U_{ab}=\mathbb{I} - i\frac{\delta\varphi}{4}\big(\mathbb{I}-\sigma_z^a-\sigma_z^b+\sigma_z^a\sigma_z^b\big),
\end{equation}  
up to $o(\delta\varphi^2)$. The resulting commutation relations write
\begin{equation}
 \sigma_k^a U_{ab}\,\delta U_{ab} = U_{ab}\,\delta U_{ab} \big(\sigma_z^b + i\frac{\delta\varphi}{2}\sigma_z^a -i\frac{\delta\varphi}{2}\sigma_z^a\sigma_z^b\big)\sigma_k^a,
\end{equation}
for $k=x,y$, while $\comm{\sigma_z^a}{U_{ab}\,\delta U_{ab}}=0$.

Yet expectation values are always real, thus only even powers of $i\delta\varphi$ can appear in their final expression.

It results that the error on the correlators computed above is at most of order $o(\delta\varphi^2)$.

\section{Conclusion}

Throughout this work, we developed a novel point of vue on GS, distinct from the stabilizer formalism, usually preferred by the community. Though the latter constitute a very powerful tool for the analysis and building of quantum algorithms on GS and many other quantum computational models, we advocate the occasional use of the somewhat more direct and intuitive tools that are correlators. 

Formally, a pure quantum state constitute a statistical distribution. As such, it is in fact fully determined by its statistical (mixed) moments. In other words, knowing all of the possible expectation values and correlators of a state amounts to knowing the whole state. 

While this turns out to be, from the computational perspective, a unreasonable way of fully characterizing a state, a number of relevant partial informations can be obtained in this way.

Correlators indeed possess the desirable property of being both easily computable and physically meaningful, deconstructing the complexity of GS in experimentally fundamental building blocks and unraveling their interaction structure.\\

Furthermore, correlators have the additional advantage of highlighting pairs of projectors that are equivalent with respect to a given state. This provides a new approach to understand the effects of projective measurements on GS, thus unravelling how multipartite entanglement emerges from mere binary interactions. It further may potentially allow for the discovery of simpler ways of implementing quantum gates.\\

An interesting follow-up of the present work would be to thoroughly examine the formal connexions between the stabilizer and correlators-based approaches, and thereby acquire an overall improved understanding of GS structures and potentialities.\\

\begin{acknowledgments}
The authors acknowledge support from the RESEARCH SUPPORT
PLAN 2022 - Call for applications for funding allocation to research projects curiosity driven (F CUR) - Project ”Entanglement Protection of Qubits’ Dynamics in a Cavity” – EPQDC , and the support by the Italian National Group of Mathematical Physics (GNFM-INdAM).
\end{acknowledgments}

\hfill

\bibliography{graph_states}

\end{document}